\documentclass[%
 reprint,
 amsmath,amssymb,
 prl,
]{revtex4-1}

\usepackage{graphicx}
\usepackage{dcolumn}
\usepackage{bm}
\usepackage{hyperref}
\usepackage{color,soul} 
\usepackage{changes}

\begin{document}

\preprint{APS/123-QED}

\title{Fundamental Limits on Measuring the Rotational Constraint of Single Molecules using Fluorescence Microscopy}

\author{Oumeng Zhang}
\author{Matthew D. Lew}%
 \email{mdlew@wustl.edu}
\affiliation{%
 Department of Electrical and Systems Engineering, Washington University in St.\ Louis, Missouri 63130, USA
}%

\date{\today}

\begin{abstract}
Optical fluorescence imaging is capable of measuring both the translational and rotational dynamics of single molecules. However, unavoidable measurement noise will result in inaccurate estimates of rotational dynamics, causing a molecule to appear to be more rotationally constrained than it actually is. We report a mathematical framework to compute the fundamental limit of accuracy in measuring the rotational mobility of dipole-like emitters. By applying our framework to both in-plane and three-dimensional methods, we provide a means to choose the optimal orientation-measurement technique based on experimental conditions.
\end{abstract}

\pacs{Valid PACS appear here}
\maketitle


Fluorescence microscopy has been widely utilized to study the rotational dynamics of single molecules (SMs), revealing new insights into DNA organization \cite{Ha1996SingleModulation,Ha1998HinderedMolecules,Backer2016EnhancedMeasurements} and the movement of molecular motors \cite{Sosa2001ADP-inducedMicroscopy,Peterman2001PolarizedMicrotubules,Forkey2003Three-dimensionalPolarization,Beausang2013TiltingMicroscopy,Lippert2017AngularStalk}. Orientation measurements are also critical for ensuring the accuracy of SM localization microscopy \cite{Engelhardt2011MolecularMicroscopy,Backlund2012SimultaneousMolecules,Lew2013RotationalMicroscopy,Lew2014AzimuthalMicroscopy}, since changes in orientation can be mistakenly perceived as changes in molecular position. To sensitively measure dipole orientation, a fluorescence microscope is typically augmented either by measuring fluorescence emission under varying pumping polarization, by manipulating the polarization and/or angular spectrum of the fluorescence emission, or both \cite{Backlund2014TheImaging}. Analyzing an SM's fluorescence signal thereby yields its average orientation and/or its rotational mobility or ``wobble" during some integration time.

One intuitive method to measure SM orientation is to quantify linear dichroism (LD) \cite{Edmiston1996DipoleAnisotropy,Fourkas2001RapidMolecules,Benninger2005FluorescenceMembranes,Steinbach2008ImagingMicroscope}, i.e., the ratio of the difference over the sum of the intensity of the $x$- and $y$-polarized emission from a single emitter. A rotationally-fixed SM will yield a certain LD value depending upon its orientation, but due to symmetry, a rotationally-unconstrained SM will give an LD of zero. However, measurement noise, e.g., photon shot noise, will almost certainly produce a nonzero LD and therefore reduce the apparent molecular wobble. This phenomenon is similar to the effect of finite localization precision in single-particle tracking \cite{Berglund2010StatisticsTracking,Wong2011LimitMicroscopy,Michalet2012OptimalTracking,Backlund2015ChromosomalErrors,Calderon2016MotionTrajectory}, where a translationally-fixed particle appears to have a nonzero diffusion coefficient due to shot noise. However, in the case of rotational dynamics, the non-zero estimates of LD cause the measurement of rotational mobility to be biased; the molecule will appear to be \emph{more constrained} than it actually is. To our knowledge, the effects of noise and measurement sensitivity and the impact of 2D versus 3D measurements on the accuracy of quantifying SM rotational dynamics remain unexplored.

Here, we present a mathematical framework to compute the distribution, accuracy, and precision of measuring the rotational constraint (inversely proportional to rotational mobility) of dipole-like emitters in both 2D and 3D. For a given imaging scenario with a certain number of signal and background photons detected, we derive a lower bound on the apparent rotational constraint, that is, the expectation of estimated rotational constraint for a freely-rotating molecule; it is impossible to detect a constraint smaller than this limit.  Further, we derive a relation between the measured in ($xy$)-plane and 3D rotational constraints, which reveals how 2D and 3D methods perceive the same 3D rotational diffusion differently. We then analyze the accuracy and precision of various commonly-used and state-of-the-art 2D and 3D orientation measurement techniques.

We model a fluorescent molecule as a radiating dipole \cite{novotny2012principles} with orientation $\boldsymbol{\mu}=[\mu_x,\mu_y,\mu_z]^\dagger$, where $\mu_z$ is the out-of-plane component, i.e., the projection of $\boldsymbol{\mu}$ along the optical axis. Since fluorescence intensity and not electric field is detected, both excitation and emission modulation methods are limited to measuring the even-order moments of $\boldsymbol{\mu}$. The recorded image $\boldsymbol{g}\in\mathbb{R}_{\geq0}^{n\times1}$, where the dimension $n$ is the number of pixels  or groupings thereof, can then be represented using the second moments of $\boldsymbol{\mu}$, given by
\begin{multline}
    \boldsymbol{g}=s[\boldsymbol{B}_{xx},\boldsymbol{B}_{yy},\boldsymbol{B}_{zz},\boldsymbol{B}_{xy},\boldsymbol{B}_{xz},\boldsymbol{B}_{yz}] \\
    [\langle\mu_x^2\rangle,\langle\mu_y^2\rangle,\langle\mu_z^2\rangle,\langle\mu_x\mu_y\rangle,\langle\mu_x\mu_z\rangle,\langle\mu_y\mu_z\rangle]^\dagger+\boldsymbol{b}
    \label{eqn:g3D}
\end{multline}
for a 3D measurement and
\begin{equation}
    \boldsymbol{g}=s[\boldsymbol{B}_{xx},\boldsymbol{B}_{yy},\boldsymbol{B}_{xy}][\langle\zeta_x^2\rangle,\langle\zeta_y^2\rangle,\langle\zeta_x\zeta_y\rangle]^\dagger+\boldsymbol{b}
    \label{eqn:g2D}
\end{equation}
for an in-plane (2D) measurement, where $\boldsymbol{B}_{ij}\in\mathbb{R}^{n\times1}$ are termed the basis images of the imaging system. The image produced by an emitter will be a mixture of these images weighted by the orientational second moments $\langle\mu_i\mu_j\rangle$. The prefactor $s$ and image $\boldsymbol{b}\in\mathbb{R}^{n\times1}$ represent the integrated signal photons and background photons, respectively, for each measurement $\boldsymbol{g}$. The second moment vector $\boldsymbol{\zeta}=[\zeta_x,\zeta_y]^\dagger=[\mu_x,\mu_y]^\dagger/\sqrt{\mu_x^2+\mu_y^2}$ is the normalized projection of molecular orientation $\boldsymbol{\mu}$ into the $xy$ plane, and the angle brackets $\langle\cdot\rangle$ denote the temporal average taken over one camera frame or the equivalent average over the orientation domain. The basis images can be computed for any excitation-modulation method by computing the SM's response to varying excitation polarization \cite{Backer2016EnhancedMeasurements} and for any emission-modulation technique using vectorial diffraction models \cite{Bohmer2003OrientationMicroscopy,Lieb2004Single-moleculeImaging,Axelrod2012FluorescenceStudy,Backer2014ExtendingProcessing,Backer2015DeterminingStudy}. 

The accuracy and precision with which the rotational constraint of a single molecule can be estimated depend on three factors: 1) the detected photons from the emitter, i.e., its signal, 2) background fluorescence, and 3) most importantly, the basis-image matrix that describes the sensitivity of the imaging system to each orientational second moment $\langle\mu_i\mu_j\rangle$. To illustrate these effects, we evaluate a simple 2D method where we pump an SM using three distinct in-plane excitation polarizations $\boldsymbol{E}_i$ and observe the corresponding emission intensities $\boldsymbol{g}=[g_1,g_2,g_3]^\dagger$ \cite{SI} (Fig.~S1(b)). 

 The normalized second moments are computed by inverting the basis matrix: $[\langle\zeta_x^2\rangle,\langle\zeta_y^2\rangle,\langle\zeta_x\zeta_y\rangle]^\dagger=\boldsymbol{B}_\text{ExMod}^{-1}(\boldsymbol{g}-\boldsymbol{b})/s$. Rearranging them into a 2-by-2 Hermitian matrix $\boldsymbol{M}_\text{2D}$ and decomposing it as
 \begin{equation}
     \boldsymbol{M}_\text{2D} = \begin{bmatrix}\langle\zeta_x^2\rangle&\langle\zeta_x\zeta_y\rangle\\
     \langle\zeta_x\zeta_y\rangle&\langle\zeta_y^2\rangle\end{bmatrix}
     = \gamma_{2D}\boldsymbol{\nu}_1\boldsymbol{\nu}_1^\dagger + \frac{1-\gamma_\text{2D}}{2}\boldsymbol{I},
     \label{eqn:gamma2D}
 \end{equation}
 we obtain the in-plane rotational constraint $\gamma_\text{2D}=2\lambda_1-1$ (Fig.~S3). The scalars $\lambda_i$ and vectors $\boldsymbol{\nu}_i$ represent the $i$th eigenvalue and eigenvector of $\boldsymbol{M}_\text{2D}$, respectively. The matrix $\boldsymbol{M}_\text{2D}$ can be interpreted as a mixture of a fixed dipole with orientation $\boldsymbol{\nu}_1$ and intensity fraction $\gamma_\text{2D}$ and an isotropic emitter with intensity fraction $1-\gamma_\text{2D}$. The rotational constraint $\gamma_\text{2D}=1$ represents a completely immobile emitter, and $\gamma_\text{2D}=0$ represents a freely-rotating molecule. Further, since rotational correlation times are typically 1-30 ns for fluorophores in liquids \cite{valeur2012molecular,Lew2013RotationalMicroscopy} and practical camera integration times are 1 ms or longer, rotational constraint measurements are typically indicative of the full range orientations explored by each molecule \cite{SI}.
 
\begin{figure}[ht]
     \centering
     \includegraphics[width=1\linewidth]{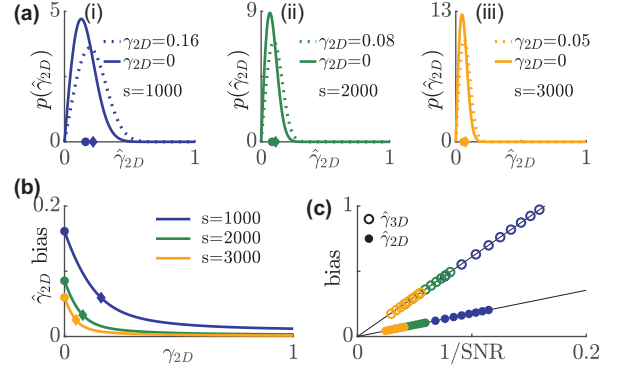}
     \caption{Distribution and bias of rotational constraint measurements $\hat{\gamma}_\text{2D}$ using in-plane excitation polarization modulation. (a) A non-central chi PDF describes the distribution of $\hat{\gamma}_\text{2D}$ under $s=$ (i) 1000, (ii) 2000, and (iii) 3000 signal photons with $\boldsymbol{1}^\dagger\boldsymbol{b}=7290$ background photons per $526.5\times526.5$~$\text{nm}^2$ (3.5$\times$ FWHM of a diffraction-limited spot) region. Solid line: isotropic emitters, dashed line: emitter whose bias is smaller by $1/e$ compared to an isotropic emitter. Circles and diamonds on the $x$ axis represent the mean of the aforementioned two distributions, respectively. (b) Average bias in $\hat{\gamma}_\text{2D}$ versus the ground truth $\gamma_\text{2D}$. Blue, green, and orange represent 1000, 2000, and 3000 signal photons, respectively. Circles and diamonds correspond to the same data points in (a). (c)~Bias in the measured rotational constraint of isotropic emitters scales linearly with the inverse of SNR, $\sqrt{s+\boldsymbol{1}^\dagger\boldsymbol{b}}/s$. Solid circles: $\hat{\gamma}_\text{2D}$, open circles: $\hat{\gamma}_\text{3D}$.} 
     \label{fig:1}
 \end{figure}

 We assume that the measured second-order moments $\hat{\langle\zeta_i\zeta_j\rangle}$ follow unbiased Gaussian distributions with precision achieving the Cram\'er-Rao lower bound (CRLB) \cite{moon2000mathematical}. Therefore, for a simplified case where the three excitation polarizations are symmetric and linear and the average orientation of the emitter lies within the $xz$ plane ($\bar{\zeta}_y=0$), the probability density function (PDF) of measurements of rotational constraint $\hat{\gamma}_\text{2D}$
  is given by \cite{SI}
 \begin{equation}
    p(\hat{\gamma}_\text{2D}) = \frac{\hat{\gamma}_\text{2D}}{4\sigma^2} \exp\left(-\frac{\hat{\gamma}_\text{2D}^2+\gamma_\text{2D}^2}{8\sigma^2}\right) I_0\left(\frac{\hat{\gamma}_\text{2D}\gamma_\text{2D}}{4\sigma^2}\right).
    \label{eqn:p_gamma2D}
 \end{equation}
 The measurement precision of the second moments is $\sigma=\sigma_{\hat{\langle\zeta_x^2\rangle}}^\text{CRLB}=\sigma_{\hat{\langle\zeta_x\zeta_y\rangle}}^\text{CRLB}$, $\gamma_\text{2D}$ is the true rotational constraint of the SM, and $I_k(\cdot)$ is the modified Bessel function of the first kind. This PDF matches simulated measurements using realistic noise (Fig.~S4). The expectation of $\hat{\gamma}_\text{2D}$ (Fig.~\ref{fig:1}(b)) is given by \cite{Nuttall1975SomeCorresp.}
 \begin{equation}
     E(\hat{\gamma}_\text{2D}) = \sigma\sqrt{2\pi} L_{1/2}^{(0)}\left(-\frac{\gamma_\text{2D}^2}{8\sigma^2}\right),
    \label{eqn:E_gamma2D}
 \end{equation}
 where $L_p^{(k)}$ is the generalized Laguerre function. Critically, this expectation does not equal the true constraint $\gamma_\text{2D}$ for any nonzero measurement precision $\sigma$.
 
 For a typical background of 30 photons per $58.5\times58.5$~$\text{nm}^2$ (one camera pixel in object space), the expected biases $\hat{\gamma}_\text{2D}-\gamma_\text{2D}$ for an isotropic emitter when 1000, 2000, 3000 signal photons are detected are 0.16, 0.09, and 0.06, respectively. To provide physical intuition, these biases correspond to half-angle errors of $13^{\circ}$, $7^{\circ}$, and $5^{\circ}$, respectively, if the molecule is uniformly diffusing within a wedge in the $xy$ plane. The error in rotational constraint decays as wobble decreases. Brighter emitters further reduce errors in rotational constraint. The bias is reduced to $1/e$ times its maximum, which occurs at $\gamma_\text{2D}=0$, when $\gamma_\text{2D}=0.16, 0.08,$ and 0.05 for 1000, 2000, and 3000 photons detected, respectively (Fig.~\ref{fig:1}(a),(b)).
 
 For an isotropic emitter, $g_i$ is invariant for different excitation polarizations, and therefore, the precision of the second-moment estimates is expressed as $\sigma=\sqrt{(s+\boldsymbol{1}^\dagger\boldsymbol{b})/2}/s$ \cite{SI}. Therefore, the average apparent rotational constraint is given by
 \begin{equation}
     E(\hat{\gamma}_\text{2D,iso})=\sqrt{\pi}\frac{\sqrt{s+\boldsymbol{1}^\dagger\boldsymbol{b}}}{s}=\frac{\sqrt{\pi}}{\mathit{SNR}},
    \label{eqn:p_gamma2Diso}
 \end{equation}
 where the bias in the apparent rotational constraint scales linearly with the inverse of the signal-to-noise ratio (SNR, Fig.~\ref{fig:1}(c)).
 
 We now extend our framework to measurements of 3D orientation, similarly assembling a $3 \times 3$ second-moment matrix and decomposing it as \cite{SI}
 \begin{equation}
     \boldsymbol{M}_\text{3D} 
     = \gamma_\text{3D}\boldsymbol{\nu}_1\boldsymbol{\nu}_1^\dagger
     +\frac{(1-\gamma_\text{3D})}{3}\boldsymbol{I}+\sum_{i=2}^3\left(\frac{(-1)^i}{2}\lambda_i\boldsymbol{\nu}_i\boldsymbol{\nu}_i^\dagger\right),
    \label{eqn:gamma3D}
 \end{equation}
 where the 3D rotational constraint $\gamma_\text{3D}=(3\lambda_1-1)/2$. If the molecule's rotational diffusion is symmetric around a certain average orientation, then the smaller eigenvalues $\lambda_2$ and $\lambda_3$ are identical due to symmetry. Similarly, the matrix $\boldsymbol{M}_\text{3D}$ can be viewed as a mixture of a fixed dipole and an isotropic emitter, plus a nuisance term that is orthogonal to $\boldsymbol{\nu}_1$ that arises from asymmetric rotation. In contrast to the case of 2D excitation modulation, the eigenvalues of the measured second-moment matrix $\hat{\boldsymbol{M}}_\text{3D}$ do not have a closed-form distribution. We therefore perform Monte Carlo simulations on isotropic emitters imaged using the Tri-spot point spread function (PSF) \cite{Zhang2018ImagingFunction} (Fig.~S1(c)). The linear relation between the bias in $\hat{\gamma}_\text{3D}$ and the inverse of SNR still approximately holds (Fig.~\ref{fig:1}(c)). 
 
 Although both 2D (Eq.~(\ref{eqn:gamma2D})) and 3D (Eq.~(\ref{eqn:gamma3D})) rotational constraints have identical interpretations for limiting cases, e.g., $\gamma_\text{2D}=\gamma_\text{3D}=0$ represents a rotationally-free emitter, these two quantities \emph{may differ} for any \emph{partially-constrained} dipole emitter. That is, identical 3D orientation trajectories can produce different rotational diffusion measurements in 2D versus 3D. Here, we consider a symmetrically-rotating molecule in 3D and assemble $\boldsymbol{M}_\text{2D}$ using a subset of elements from $\boldsymbol{M}_\text{3D}$ \cite{SI}. The in-plane rotational constraint $\gamma_\text{2D}$ is now given by
 \begin{equation}
    \gamma_\text{2D} = \frac{3(1-\bar{\mu}_z^2)}
    {1-3\bar{\mu}_z^2+2/\gamma_\text{3D}}.
    \label{eqn:gamma2D_3D}
 \end{equation}
 
 \begin{figure}[ht]
    \centering
    \includegraphics[width=1\linewidth]{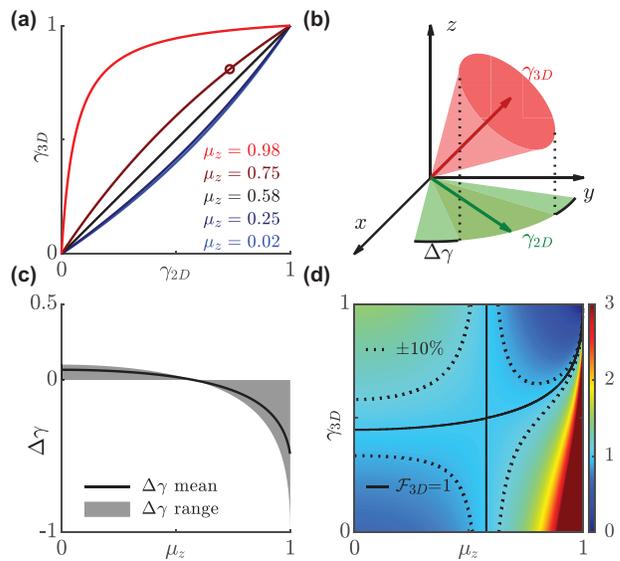}
    \caption{Comparison of in-plane $\gamma_\text{2D}$ and 3D $\gamma_\text{3D}$ rotational constraints. (a) The relation between $\gamma_\text{2D}$ and $\gamma_\text{3D}$ varies with the average orientation along the $z$ axis, $\bar{\mu}_z$ (Eq.~\ref{eqn:gamma2D_3D}). Black line corresponds to $\bar{\mu}_z = \sqrt{1/3} \approx \cos 54.7^\circ$ so that $\gamma_\text{2D}=\gamma_\text{3D}$. (b) A dipole emitter with an out-of-plane orientation $\bar{\mu}_z=0.75$ that has $\gamma_\text{3D}=0.81$ (a cone half-angle of $30^\circ$, open circle in (a)) appears to be more rotationally free in the $xy$ plane ($\gamma_\text{2D}=0.73$, wedge half-angle of $38^\circ$). (c) The mean difference $\Delta\gamma$ across all 3D orientation space ranges from -1 to 0.06. Solid line represents the average across all possible values of $\gamma_\text{3D}$, and the shaded region represents the range of the difference. (d) 3D enhancement factor $\mathcal{F}_\text{3D}$ as a function of $\gamma_\text{3D}$ and $\bar{\mu}_z$. Solid line represents the equilibrium point where in-plane and 3D rotational constraints measure changes in rotational dynamics with equal sensitivity. Dashed line represents a $\pm10\%$ difference in sensitivity.}
    \label{fig:2}
 \end{figure}
 
 The 2D and 3D rotational constraints are identical for a molecule exhibiting an average out-of-plane component of $\bar{\mu}_z^2=1/3$ (corresponding to a ``magic" polar angle $\approx54.7^\circ$, Fig.~\ref{fig:2}a). Note that $\gamma_\text{2D}$ depends on \emph{both} the 3D rotational constraint $\gamma_\text{3D}$ and average out-of-plane orientation $\bar{\mu}_z$; in-plane orientation measurement methods are only sensitive to $\gamma_\text{2D}$ and must incorporate prior knowledge of $\bar{\mu}_z$ in order to compute an equivalent 3D rotational constraint $\gamma_\text{3D}$. The average difference in constraint $\Delta\gamma=\gamma_\text{2D}-\gamma_\text{3D}$ (Fig.~\ref{fig:2}(c)) is within $\pm0.1$ for $\bar{\mu}_z\leq0.8$, indicating that both $\gamma_\text{2D}$ and $\gamma_\text{3D}$ quantify the rotational dynamics of a dipole emitter similarly as long as the out-of-plane component is small. However, for a molecule that is almost along the optical axis, e.g. $\bar{\mu}_z=0.98$ or polar angle = $11^\circ$, $|\Delta\gamma|$ can be as large as 0.61; a highly-constrained molecule in 3D ($\gamma_\text{3D}=0.80$ or a cone half-angle of $30^\circ$) appears to be almost completely unconstrained using an in-plane measurement method ($\gamma_\text{2D}=0.20$ or a wedge half-angle of $75^\circ$). This interdependence of $\mu_z$, $\gamma_\text{2D}$, and $\gamma_\text{3D}$ has important implications for orientation-measurement techniques. If a technique cannot measure all six second-moments in 3D directly, then one must use a prior assumption on the out-of-plane orientation to calculate the rotational constraint (or vice versa). Any errors in this assumption can dramatically impact measurement accuracy. 
 
 A natural consequence of using 2D versus 3D orientation measurements is that these techniques have different sensitivities for measuring changes in rotational dynamics. Here, we quantify the enhancement factor $\mathcal{F}_\text{3D}$ as the ratio of partial derivatives of $\gamma_\text{3D}$ to $\gamma_\text{2D}$ (Fig.~\ref{fig:2}(d)):
 \begin{equation}
     \mathcal{F}_\text{3D} = \frac{\partial\gamma_\text{3D}}{\partial\gamma_\text{2D}} = \frac{\left(\gamma_\text{3D}(1-3\bar{\mu}_z^2)+2\right)^2}{6(1-\bar{\mu}_z^2)}.
    \label{eqn:f3D}
 \end{equation}
 An enhancement factor $\mathcal{F}_\text{3D}$ greater than one implies that a given change in $\hat{\gamma}_\text{3D}$ maps to a smaller change in $\hat{\gamma}_\text{2D}$, i.e., for the same measurement uncertainty, it is easier to detect a change in $\hat{\gamma}_\text{3D}$ than $\hat{\gamma}_\text{2D}$. The sensitivity of the in-plane measurement to a change in rotational constraint highly depends on the out-of-plane component of molecular orientation. For example, for the aforementioned molecule with $\bar{\mu}_z=0.98$, an in-plane technique is especially insensitive to changes in rotational motion for most values of $\gamma_\text{3D}$. However, when the molecule is almost immobile, the in-plane method becomes very sensitive, i.e., a small change in $\gamma_\text{3D}$ produces a large, easily-detectable change in $\hat{\gamma}_\text{2D}$.
 
  \begin{figure}[ht]
     \centering
     \includegraphics[width=1\linewidth]{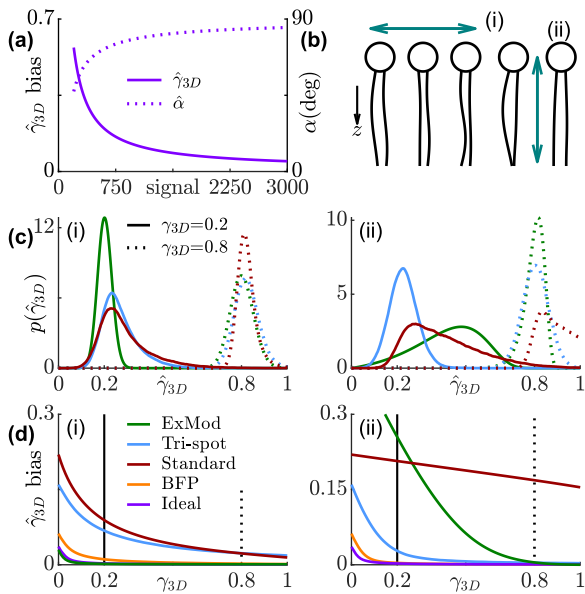}
     \caption{Bias and precision of rotational constraint measurements using various techniques. (a) Minimum detectable rotational constraint for various numbers of signal photons and 4860 total background photons. Solid line represents the expectation of measured rotational constraint $\hat{\gamma}_\text{3D}$ for isotropic emitters; dashed line represents the corresponding cone half-angle $\hat{\alpha}$ if the molecule is symmetrically diffusing in a cone. (b) Concept of (i) a horizontally- and (ii) a vertically-orientated fluorescent molecule within a lipid membrane. (c)~Distribution of 3D rotational constraint estimates $\hat{\gamma}_\text{3D}$. Solid line: true rotational constraint $\gamma_\text{3D}=0.2$ (cone half-angle $\alpha=72^\circ$); dashed line: $\gamma_\text{3D}=0.8$ ($\alpha=31^\circ$). (d) Bias of $\hat{\gamma}_\text{3D}$ for each technique. Green: in-plane excitation modulation, blue: Tri-spot PSF, red: standard PSF, orange: back focal plane imaging, purple: ideal basis-image matrix.}
     \label{fig:3}
 \end{figure}
 
 To provide quantitative metrics for choosing the optimal orientation measurement technique, we compare multiple popular and state-of-the-art methods. We first consider an ideal 3D ($6\times6$) basis-image matrix \cite{SI}
 \begin{equation}
    \boldsymbol{B}_{\text{ideal}} =\frac{1}{2} \begin{bmatrix}\boldsymbol{I}_3 & \boldsymbol{I}_3\\
    \boldsymbol{I}_3 & -\boldsymbol{I}_3\end{bmatrix},
    \label{eqn:bideal}
 \end{equation}
 such that each column of $\boldsymbol{B}_\text{ideal}$ is orthogonal and has identical energy. The average apparent 3D rotational constraint for an isotropic emitter is given by \cite{SI}
 \begin{equation}
    E(\hat{\gamma}_\text{3D,iso}) = \frac{\eta\sqrt{3}}{2}\frac{\sqrt{s+\boldsymbol{1}^\dagger\boldsymbol{b}}}{s}=\frac{\eta\sqrt{3}}{2\,\mathit{SNR}},
    \label{eqn:Egamma3Diso}
 \end{equation}
 where $\eta\approx1.848$ is the expectation of the largest eigenvalue of a Hermitian random matrix $\boldsymbol{H}$ whose elements $H_{ij}$ are i.i.d.\ standard normal random variables. The imaging system represented by $\boldsymbol{B}_\text{ideal}$ separates the six second-moments into independent measurements with ideal collection efficiency and, thus, represents a fundamental limit on the measurement accuracy of 3D rotational constraint. That is, for a finite SNR, a freely-rotating molecule would appear to be identical to a partially-fixed emitter using any 3D orientation measurement method, and the apparent constraint is always greater than or equal to that measured by Eq.~(\ref{eqn:Egamma3Diso}) \cite{SI}. To provide physical intuition (Fig.~\ref{fig:3}(a)), for 1000 signal photons and 30 background photons per pixel detected, the expectation of $\hat{\gamma}_\text{3D}$ is 0.12, meaning that an SM diffusing within a cone of half-angle $\alpha=78^\circ$ would be indistinguishable from a rotationally-free SM ($\alpha=90^\circ$).
 
 Besides the aforementioned ideal, Tri-spot, and in-plane excitation methods, we also analyze the performance of fitting fine features of the standard PSF \cite{Mortensen2010OptimizedMicroscopy} and direct imaging of the back focal plane (BFP) \cite{Lieb2004Single-moleculeImaging} (Fig.~S1(d,e)). We simulated two limiting cases, mimicking two orientations of fluorescent dye molecules embedded within a lipid membrane \cite{Motegi2013Single-moleculeBilayers} (Fig.~\ref{fig:3}(b)), where the average orientation is almost perpendicular (Fig.~\ref{fig:3}(i), $\bar{\mu}_z=0.02$) and almost parallel (Fig.~\ref{fig:3}(ii), $\bar{\mu}_z=0.98$) to the optical axis. The simulated SNR for all the methods is 3000 signal photons to 20 total background photons per pixel per unit time \cite{SI}. 
 
 The Tri-spot PSF and BFP imaging exhibit consistent performance for both in-plane and out-of-plane molecules due to their relatively uniform sensitivity for measuring all orientational second-order moments. In-plane excitation, as expected, exhibits high accuracy and precision (Fig. S7) for in-plane molecules. Its bias in $\hat{\gamma}_\text{3D}$ is even lower than the ideal 3D technique because of the improved SNR of distributing an SM's photons over 2$\times$ fewer measurements. For out-of-plane molecules however, its bias and precision are much worse compared to 3D methods except when the rotational constraint is sufficiently high. The standard PSF, due to its poor sensitivity to two out-of-plane second moments, also exhibits a large bias and standard deviation for out-of-plane molecules. However, it has better performance for measuring the rotational motion of highly constrained in-plane molecules than the Tri-spot PSF. This analysis implies that a method specifically designed for measuring a subset of, instead of all six (3D), orientational second moments can provide superior performance under certain experimental conditions.
 
 In summary, we analyzed the fundamental bias in measurements of rotational dynamics caused by finite SNR. We derived a lower bound on this bias, that is, no method can detect a rotational constraint smaller than this bound without sufficient prior knowledge. While we defined rotational constraint in terms of the eigenvalues of the second-moment matrix $\boldsymbol{M}$, any parameterization of rotational diffusion, e.g., fluorescence anisotropy \cite{lakowicz2004principles,SI}, or choice of emitter model, e.g., higher-order multipoles \cite{novotny2012principles,SI}, will also suffer bias since any measurement will capture a finite number of photons. Our framework is easily adaptable for characterizing the bias of any orientation-measurement method via calculation of the basis images $\boldsymbol{B}_{ij}$. We show that there is a complex relationship between in-plane and 3D molecular orientation, such that a molecule's 2D rotational constraint can appear significantly smaller or larger than its true constraint in 3D. Using our framework, we compared multiple methods for measuring rotational dynamics, revealing how the accuracy and precision of these measurements vary for in-plane and out-of-plane SMs. In particular, our results show that 3D methods are particularly important for quantifying accurately the rotational dynamics of SMs with significant out-of-plane orientations. Our framework should be useful for choosing between existing methods and optimizing new techniques that achieve maximum accuracy and precision in various imaging scenarios.
 
 \begin{acknowledgments}
 This work was supported by the National Science Foundation under grant number ECCS-1653777 and by the National Institute of General Medical Sciences of the National Institutes of Health under grant number R35GM124858.
 \end{acknowledgments}

\bibliographystyle{apsrev4-1}
\bibliography{references}

\begin{thebibliography}{41}%
\makeatletter
\providecommand \@ifxundefined [1]{%
 \@ifx{#1\undefined}
}%
\providecommand \@ifnum [1]{%
 \ifnum #1\expandafter \@firstoftwo
 \else \expandafter \@secondoftwo
 \fi
}%
\providecommand \@ifx [1]{%
 \ifx #1\expandafter \@firstoftwo
 \else \expandafter \@secondoftwo
 \fi
}%
\providecommand \natexlab [1]{#1}%
\providecommand \enquote  [1]{``#1''}%
\providecommand \bibnamefont  [1]{#1}%
\providecommand \bibfnamefont [1]{#1}%
\providecommand \citenamefont [1]{#1}%
\providecommand \href@noop [0]{\@secondoftwo}%
\providecommand \href [0]{\begingroup \@sanitize@url \@href}%
\providecommand \@href[1]{\@@startlink{#1}\@@href}%
\providecommand \@@href[1]{\endgroup#1\@@endlink}%
\providecommand \@sanitize@url [0]{\catcode `\\12\catcode `\$12\catcode
  `\&12\catcode `\#12\catcode `\^12\catcode `\_12\catcode `\%12\relax}%
\providecommand \@@startlink[1]{}%
\providecommand \@@endlink[0]{}%
\providecommand \url  [0]{\begingroup\@sanitize@url \@url }%
\providecommand \@url [1]{\endgroup\@href {#1}{\urlprefix }}%
\providecommand \urlprefix  [0]{URL }%
\providecommand \Eprint [0]{\href }%
\providecommand \doibase [0]{http://dx.doi.org/}%
\providecommand \selectlanguage [0]{\@gobble}%
\providecommand \bibinfo  [0]{\@secondoftwo}%
\providecommand \bibfield  [0]{\@secondoftwo}%
\providecommand \translation [1]{[#1]}%
\providecommand \BibitemOpen [0]{}%
\providecommand \bibitemStop [0]{}%
\providecommand \bibitemNoStop [0]{.\EOS\space}%
\providecommand \EOS [0]{\spacefactor3000\relax}%
\providecommand \BibitemShut  [1]{\csname bibitem#1\endcsname}%
\let\auto@bib@innerbib\@empty
\bibitem [{\citenamefont {Ha}\ \emph {et~al.}(1996)\citenamefont {Ha},
  \citenamefont {Enderle}, \citenamefont {Chemla}, \citenamefont {Selvin},\
  and\ \citenamefont {Weiss}}]{Ha1996SingleModulation}%
  \BibitemOpen
  \bibfield  {author} {\bibinfo {author} {\bibfnamefont {T.}~\bibnamefont
  {Ha}}, \bibinfo {author} {\bibfnamefont {T.}~\bibnamefont {Enderle}},
  \bibinfo {author} {\bibfnamefont {D.~S.}\ \bibnamefont {Chemla}}, \bibinfo
  {author} {\bibfnamefont {P.~R.}\ \bibnamefont {Selvin}}, \ and\ \bibinfo
  {author} {\bibfnamefont {S.}~\bibnamefont {Weiss}},\ }\href {\doibase
  10.1103/PhysRevLett.77.3979} {\bibfield  {journal} {\bibinfo  {journal}
  {Physical Review Letters}\ }\textbf {\bibinfo {volume} {77}},\ \bibinfo
  {pages} {3979} (\bibinfo {year} {1996})}\BibitemShut {NoStop}%
\bibitem [{\citenamefont {Ha}\ \emph {et~al.}(1998)\citenamefont {Ha},
  \citenamefont {Glass}, \citenamefont {Enderle}, \citenamefont {Chemla},\ and\
  \citenamefont {Weiss}}]{Ha1998HinderedMolecules}%
  \BibitemOpen
  \bibfield  {author} {\bibinfo {author} {\bibfnamefont {T.}~\bibnamefont
  {Ha}}, \bibinfo {author} {\bibfnamefont {J.}~\bibnamefont {Glass}}, \bibinfo
  {author} {\bibfnamefont {T.}~\bibnamefont {Enderle}}, \bibinfo {author}
  {\bibfnamefont {D.~S.}\ \bibnamefont {Chemla}}, \ and\ \bibinfo {author}
  {\bibfnamefont {S.}~\bibnamefont {Weiss}},\ }\href {\doibase
  10.1103/PhysRevLett.80.2093} {\bibfield  {journal} {\bibinfo  {journal}
  {Physical Review Letters}\ }\textbf {\bibinfo {volume} {80}},\ \bibinfo
  {pages} {2093} (\bibinfo {year} {1998})}\BibitemShut {NoStop}%
\bibitem [{\citenamefont {Backer}\ \emph {et~al.}(2016)\citenamefont {Backer},
  \citenamefont {Lee},\ and\ \citenamefont
  {Moerner}}]{Backer2016EnhancedMeasurements}%
  \BibitemOpen
  \bibfield  {author} {\bibinfo {author} {\bibfnamefont {A.~S.}\ \bibnamefont
  {Backer}}, \bibinfo {author} {\bibfnamefont {M.~Y.}\ \bibnamefont {Lee}}, \
  and\ \bibinfo {author} {\bibfnamefont {W.~E.}\ \bibnamefont {Moerner}},\
  }\href {\doibase 10.1364/OPTICA.3.000659} {\bibfield  {journal} {\bibinfo
  {journal} {Optica}\ }\textbf {\bibinfo {volume} {3}},\ \bibinfo {pages} {659}
  (\bibinfo {year} {2016})}\BibitemShut {NoStop}%
\bibitem [{\citenamefont {Sosa}\ \emph {et~al.}(2001)\citenamefont {Sosa},
  \citenamefont {Peterman}, \citenamefont {Moerner},\ and\ \citenamefont
  {Goldstein}}]{Sosa2001ADP-inducedMicroscopy}%
  \BibitemOpen
  \bibfield  {author} {\bibinfo {author} {\bibfnamefont {H.}~\bibnamefont
  {Sosa}}, \bibinfo {author} {\bibfnamefont {E.~J.~G.}\ \bibnamefont
  {Peterman}}, \bibinfo {author} {\bibfnamefont {W.~E.}\ \bibnamefont
  {Moerner}}, \ and\ \bibinfo {author} {\bibfnamefont {L.~S.~B.}\ \bibnamefont
  {Goldstein}},\ }\href {\doibase 10.1038/88611} {\bibfield  {journal}
  {\bibinfo  {journal} {Nature Structural Biology}\ }\textbf {\bibinfo {volume}
  {8}},\ \bibinfo {pages} {540} (\bibinfo {year} {2001})}\BibitemShut {NoStop}%
\bibitem [{\citenamefont {Peterman}\ \emph {et~al.}(2001)\citenamefont
  {Peterman}, \citenamefont {Sosa}, \citenamefont {Goldstein},\ and\
  \citenamefont {Moerner}}]{Peterman2001PolarizedMicrotubules}%
  \BibitemOpen
  \bibfield  {author} {\bibinfo {author} {\bibfnamefont {E.~J.}\ \bibnamefont
  {Peterman}}, \bibinfo {author} {\bibfnamefont {H.}~\bibnamefont {Sosa}},
  \bibinfo {author} {\bibfnamefont {L.~S.}\ \bibnamefont {Goldstein}}, \ and\
  \bibinfo {author} {\bibfnamefont {W.}~\bibnamefont {Moerner}},\ }\href
  {\doibase 10.1016/S0006-3495(01)75926-7} {\bibfield  {journal} {\bibinfo
  {journal} {Biophysical Journal}\ }\textbf {\bibinfo {volume} {81}},\ \bibinfo
  {pages} {2851} (\bibinfo {year} {2001})}\BibitemShut {NoStop}%
\bibitem [{\citenamefont {Forkey}\ \emph {et~al.}(2003)\citenamefont {Forkey},
  \citenamefont {Quinlan}, \citenamefont {Shaw}, \citenamefont {Corrie},\ and\
  \citenamefont {Goldman}}]{Forkey2003Three-dimensionalPolarization}%
  \BibitemOpen
  \bibfield  {author} {\bibinfo {author} {\bibfnamefont {J.~N.}\ \bibnamefont
  {Forkey}}, \bibinfo {author} {\bibfnamefont {M.~E.}\ \bibnamefont {Quinlan}},
  \bibinfo {author} {\bibfnamefont {M.~A.}\ \bibnamefont {Shaw}}, \bibinfo
  {author} {\bibfnamefont {J.~E.~T.}\ \bibnamefont {Corrie}}, \ and\ \bibinfo
  {author} {\bibfnamefont {Y.~E.}\ \bibnamefont {Goldman}},\ }\href {\doibase
  10.1038/nature01529} {\bibfield  {journal} {\bibinfo  {journal} {Nature}\
  }\textbf {\bibinfo {volume} {422}},\ \bibinfo {pages} {399} (\bibinfo {year}
  {2003})}\BibitemShut {NoStop}%
\bibitem [{\citenamefont {Beausang}\ \emph {et~al.}(2013)\citenamefont
  {Beausang}, \citenamefont {Shroder}, \citenamefont {Nelson},\ and\
  \citenamefont {Goldman}}]{Beausang2013TiltingMicroscopy}%
  \BibitemOpen
  \bibfield  {author} {\bibinfo {author} {\bibfnamefont {J.~F.}\ \bibnamefont
  {Beausang}}, \bibinfo {author} {\bibfnamefont {D.~Y.}\ \bibnamefont
  {Shroder}}, \bibinfo {author} {\bibfnamefont {P.~C.}\ \bibnamefont {Nelson}},
  \ and\ \bibinfo {author} {\bibfnamefont {Y.~E.}\ \bibnamefont {Goldman}},\
  }\href {\doibase 10.1016/j.bpj.2013.01.057} {\bibfield  {journal} {\bibinfo
  {journal} {Biophysical Journal}\ }\textbf {\bibinfo {volume} {104}},\
  \bibinfo {pages} {1263} (\bibinfo {year} {2013})}\BibitemShut {NoStop}%
\bibitem [{\citenamefont {Lippert}\ \emph {et~al.}(2017)\citenamefont
  {Lippert}, \citenamefont {Dadosh}, \citenamefont {Hadden}, \citenamefont
  {Karnawat}, \citenamefont {Diroll}, \citenamefont {Murray}, \citenamefont
  {Holzbaur}, \citenamefont {Schulten}, \citenamefont {Reck-Peterson},\ and\
  \citenamefont {Goldman}}]{Lippert2017AngularStalk}%
  \BibitemOpen
  \bibfield  {author} {\bibinfo {author} {\bibfnamefont {L.~G.}\ \bibnamefont
  {Lippert}}, \bibinfo {author} {\bibfnamefont {T.}~\bibnamefont {Dadosh}},
  \bibinfo {author} {\bibfnamefont {J.~A.}\ \bibnamefont {Hadden}}, \bibinfo
  {author} {\bibfnamefont {V.}~\bibnamefont {Karnawat}}, \bibinfo {author}
  {\bibfnamefont {B.~T.}\ \bibnamefont {Diroll}}, \bibinfo {author}
  {\bibfnamefont {C.~B.}\ \bibnamefont {Murray}}, \bibinfo {author}
  {\bibfnamefont {E.~L.~F.}\ \bibnamefont {Holzbaur}}, \bibinfo {author}
  {\bibfnamefont {K.}~\bibnamefont {Schulten}}, \bibinfo {author}
  {\bibfnamefont {S.~L.}\ \bibnamefont {Reck-Peterson}}, \ and\ \bibinfo
  {author} {\bibfnamefont {Y.~E.}\ \bibnamefont {Goldman}},\ }\href {\doibase
  10.1073/pnas.1620149114} {\bibfield  {journal} {\bibinfo  {journal}
  {Proceedings of the National Academy of Sciences}\ }\textbf {\bibinfo
  {volume} {114}},\ \bibinfo {pages} {E4564} (\bibinfo {year}
  {2017})}\BibitemShut {NoStop}%
\bibitem [{\citenamefont {Engelhardt}\ \emph {et~al.}(2011)\citenamefont
  {Engelhardt}, \citenamefont {Keller}, \citenamefont {Hoyer}, \citenamefont
  {Reuss}, \citenamefont {Staudt},\ and\ \citenamefont
  {Hell}}]{Engelhardt2011MolecularMicroscopy}%
  \BibitemOpen
  \bibfield  {author} {\bibinfo {author} {\bibfnamefont {J.}~\bibnamefont
  {Engelhardt}}, \bibinfo {author} {\bibfnamefont {J.}~\bibnamefont {Keller}},
  \bibinfo {author} {\bibfnamefont {P.}~\bibnamefont {Hoyer}}, \bibinfo
  {author} {\bibfnamefont {M.}~\bibnamefont {Reuss}}, \bibinfo {author}
  {\bibfnamefont {T.}~\bibnamefont {Staudt}}, \ and\ \bibinfo {author}
  {\bibfnamefont {S.~W.}\ \bibnamefont {Hell}},\ }\href {\doibase
  10.1021/nl103472b} {\bibfield  {journal} {\bibinfo  {journal} {Nano Letters}\
  }\textbf {\bibinfo {volume} {11}},\ \bibinfo {pages} {209} (\bibinfo {year}
  {2011})}\BibitemShut {NoStop}%
\bibitem [{\citenamefont {Backlund}\ \emph {et~al.}(2012)\citenamefont
  {Backlund}, \citenamefont {Lew}, \citenamefont {Backer}, \citenamefont
  {Sahl}, \citenamefont {Grover}, \citenamefont {Agrawal}, \citenamefont
  {Piestun},\ and\ \citenamefont
  {Moerner}}]{Backlund2012SimultaneousMolecules}%
  \BibitemOpen
  \bibfield  {author} {\bibinfo {author} {\bibfnamefont {M.~P.}\ \bibnamefont
  {Backlund}}, \bibinfo {author} {\bibfnamefont {M.~D.}\ \bibnamefont {Lew}},
  \bibinfo {author} {\bibfnamefont {A.~S.}\ \bibnamefont {Backer}}, \bibinfo
  {author} {\bibfnamefont {S.~J.}\ \bibnamefont {Sahl}}, \bibinfo {author}
  {\bibfnamefont {G.}~\bibnamefont {Grover}}, \bibinfo {author} {\bibfnamefont
  {A.}~\bibnamefont {Agrawal}}, \bibinfo {author} {\bibfnamefont
  {R.}~\bibnamefont {Piestun}}, \ and\ \bibinfo {author} {\bibfnamefont
  {W.~E.}\ \bibnamefont {Moerner}},\ }\href {\doibase 10.1073/pnas.1216687109}
  {\bibfield  {journal} {\bibinfo  {journal} {Proceedings of the National
  Academy of Sciences}\ }\textbf {\bibinfo {volume} {109}},\ \bibinfo {pages}
  {19087} (\bibinfo {year} {2012})}\BibitemShut {NoStop}%
\bibitem [{\citenamefont {Lew}\ \emph {et~al.}(2013)\citenamefont {Lew},
  \citenamefont {Backlund},\ and\ \citenamefont
  {Moerner}}]{Lew2013RotationalMicroscopy}%
  \BibitemOpen
  \bibfield  {author} {\bibinfo {author} {\bibfnamefont {M.~D.}\ \bibnamefont
  {Lew}}, \bibinfo {author} {\bibfnamefont {M.~P.}\ \bibnamefont {Backlund}}, \
  and\ \bibinfo {author} {\bibfnamefont {W.~E.}\ \bibnamefont {Moerner}},\
  }\href {\doibase 10.1021/nl304359p} {\bibfield  {journal} {\bibinfo
  {journal} {Nano Letters}\ }\textbf {\bibinfo {volume} {13}},\ \bibinfo
  {pages} {3967} (\bibinfo {year} {2013})}\BibitemShut {NoStop}%
\bibitem [{\citenamefont {Lew}\ and\ \citenamefont
  {Moerner}(2014)}]{Lew2014AzimuthalMicroscopy}%
  \BibitemOpen
  \bibfield  {author} {\bibinfo {author} {\bibfnamefont {M.~D.}\ \bibnamefont
  {Lew}}\ and\ \bibinfo {author} {\bibfnamefont {W.~E.}\ \bibnamefont
  {Moerner}},\ }\href {\doibase 10.1021/nl502914k} {\bibfield  {journal}
  {\bibinfo  {journal} {Nano Letters}\ }\textbf {\bibinfo {volume} {14}},\
  \bibinfo {pages} {6407} (\bibinfo {year} {2014})}\BibitemShut {NoStop}%
\bibitem [{\citenamefont {Backlund}\ \emph {et~al.}(2014)\citenamefont
  {Backlund}, \citenamefont {Lew}, \citenamefont {Backer}, \citenamefont
  {Sahl},\ and\ \citenamefont {Moerner}}]{Backlund2014TheImaging}%
  \BibitemOpen
  \bibfield  {author} {\bibinfo {author} {\bibfnamefont {M.~P.}\ \bibnamefont
  {Backlund}}, \bibinfo {author} {\bibfnamefont {M.~D.}\ \bibnamefont {Lew}},
  \bibinfo {author} {\bibfnamefont {A.~S.}\ \bibnamefont {Backer}}, \bibinfo
  {author} {\bibfnamefont {S.~J.}\ \bibnamefont {Sahl}}, \ and\ \bibinfo
  {author} {\bibfnamefont {W.~E.}\ \bibnamefont {Moerner}},\ }\href {\doibase
  10.1002/cphc.201300880} {\bibfield  {journal} {\bibinfo  {journal}
  {ChemPhysChem}\ }\textbf {\bibinfo {volume} {15}},\ \bibinfo {pages} {587}
  (\bibinfo {year} {2014})}\BibitemShut {NoStop}%
\bibitem [{\citenamefont {Edmiston}\ \emph {et~al.}(1996)\citenamefont
  {Edmiston}, \citenamefont {Lee}, \citenamefont {Wood},\ and\ \citenamefont
  {Saavedra}}]{Edmiston1996DipoleAnisotropy}%
  \BibitemOpen
  \bibfield  {author} {\bibinfo {author} {\bibfnamefont {P.~L.}\ \bibnamefont
  {Edmiston}}, \bibinfo {author} {\bibfnamefont {J.~E.}\ \bibnamefont {Lee}},
  \bibinfo {author} {\bibfnamefont {L.~L.}\ \bibnamefont {Wood}}, \ and\
  \bibinfo {author} {\bibfnamefont {S.~S.}\ \bibnamefont {Saavedra}},\ }\href
  {\doibase 10.1021/jp952037p} {\bibfield  {journal} {\bibinfo  {journal} {The
  Journal of Physical Chemistry}\ }\textbf {\bibinfo {volume} {100}},\ \bibinfo
  {pages} {775} (\bibinfo {year} {1996})}\BibitemShut {NoStop}%
\bibitem [{\citenamefont {Fourkas}(2001)}]{Fourkas2001RapidMolecules}%
  \BibitemOpen
  \bibfield  {author} {\bibinfo {author} {\bibfnamefont {J.~T.}\ \bibnamefont
  {Fourkas}},\ }\href {\doibase 10.1364/OL.26.000211} {\bibfield  {journal}
  {\bibinfo  {journal} {Optics Letters}\ }\textbf {\bibinfo {volume} {26}},\
  \bibinfo {pages} {211} (\bibinfo {year} {2001})}\BibitemShut {NoStop}%
\bibitem [{\citenamefont {Benninger}\ \emph {et~al.}(2005)\citenamefont
  {Benninger}, \citenamefont {{\"{O}}nfelt}, \citenamefont {Neil},
  \citenamefont {Davis},\ and\ \citenamefont
  {French}}]{Benninger2005FluorescenceMembranes}%
  \BibitemOpen
  \bibfield  {author} {\bibinfo {author} {\bibfnamefont {R.~K.}\ \bibnamefont
  {Benninger}}, \bibinfo {author} {\bibfnamefont {B.}~\bibnamefont
  {{\"{O}}nfelt}}, \bibinfo {author} {\bibfnamefont {M.~A.}\ \bibnamefont
  {Neil}}, \bibinfo {author} {\bibfnamefont {D.~M.}\ \bibnamefont {Davis}}, \
  and\ \bibinfo {author} {\bibfnamefont {P.~M.}\ \bibnamefont {French}},\
  }\href {\doibase 10.1529/biophysj.104.050096} {\bibfield  {journal} {\bibinfo
   {journal} {Biophysical Journal}\ }\textbf {\bibinfo {volume} {88}},\
  \bibinfo {pages} {609} (\bibinfo {year} {2005})}\BibitemShut {NoStop}%
\bibitem [{\citenamefont {Steinbach}\ \emph {et~al.}(2008)\citenamefont
  {Steinbach}, \citenamefont {Pomozi}, \citenamefont {Zsiros}, \citenamefont
  {P{\'{a}}y}, \citenamefont {Horv{\'{a}}th},\ and\ \citenamefont
  {Garab}}]{Steinbach2008ImagingMicroscope}%
  \BibitemOpen
  \bibfield  {author} {\bibinfo {author} {\bibfnamefont {G.}~\bibnamefont
  {Steinbach}}, \bibinfo {author} {\bibfnamefont {I.}~\bibnamefont {Pomozi}},
  \bibinfo {author} {\bibfnamefont {O.}~\bibnamefont {Zsiros}}, \bibinfo
  {author} {\bibfnamefont {A.}~\bibnamefont {P{\'{a}}y}}, \bibinfo {author}
  {\bibfnamefont {G.~V.}\ \bibnamefont {Horv{\'{a}}th}}, \ and\ \bibinfo
  {author} {\bibfnamefont {G.}~\bibnamefont {Garab}},\ }\href {\doibase
  10.1002/cyto.a.20517} {\bibfield  {journal} {\bibinfo  {journal} {Cytometry
  Part A}\ }\textbf {\bibinfo {volume} {73A}},\ \bibinfo {pages} {202}
  (\bibinfo {year} {2008})}\BibitemShut {NoStop}%
\bibitem [{\citenamefont {Berglund}(2010)}]{Berglund2010StatisticsTracking}%
  \BibitemOpen
  \bibfield  {author} {\bibinfo {author} {\bibfnamefont {A.~J.}\ \bibnamefont
  {Berglund}},\ }\href {\doibase 10.1103/PhysRevE.82.011917} {\bibfield
  {journal} {\bibinfo  {journal} {Physical Review E}\ }\textbf {\bibinfo
  {volume} {82}},\ \bibinfo {pages} {011917} (\bibinfo {year}
  {2010})}\BibitemShut {NoStop}%
\bibitem [{\citenamefont {Wong}\ \emph {et~al.}(2011)\citenamefont {Wong},
  \citenamefont {Lin},\ and\ \citenamefont {Ober}}]{Wong2011LimitMicroscopy}%
  \BibitemOpen
  \bibfield  {author} {\bibinfo {author} {\bibfnamefont {Y.}~\bibnamefont
  {Wong}}, \bibinfo {author} {\bibfnamefont {Z.}~\bibnamefont {Lin}}, \ and\
  \bibinfo {author} {\bibfnamefont {R.~J.}\ \bibnamefont {Ober}},\ }\href
  {\doibase 10.1109/TSP.2010.2098403} {\bibfield  {journal} {\bibinfo
  {journal} {IEEE Transactions on Signal Processing}\ }\textbf {\bibinfo
  {volume} {59}},\ \bibinfo {pages} {895} (\bibinfo {year} {2011})}\BibitemShut
  {NoStop}%
\bibitem [{\citenamefont {Michalet}\ and\ \citenamefont
  {Berglund}(2012)}]{Michalet2012OptimalTracking}%
  \BibitemOpen
  \bibfield  {author} {\bibinfo {author} {\bibfnamefont {X.}~\bibnamefont
  {Michalet}}\ and\ \bibinfo {author} {\bibfnamefont {A.~J.}\ \bibnamefont
  {Berglund}},\ }\href {\doibase 10.1103/PhysRevE.85.061916} {\bibfield
  {journal} {\bibinfo  {journal} {Physical Review E}\ }\textbf {\bibinfo
  {volume} {85}},\ \bibinfo {pages} {061916} (\bibinfo {year}
  {2012})}\BibitemShut {NoStop}%
\bibitem [{\citenamefont {Backlund}\ \emph {et~al.}(2015)\citenamefont
  {Backlund}, \citenamefont {Joyner},\ and\ \citenamefont
  {Moerner}}]{Backlund2015ChromosomalErrors}%
  \BibitemOpen
  \bibfield  {author} {\bibinfo {author} {\bibfnamefont {M.~P.}\ \bibnamefont
  {Backlund}}, \bibinfo {author} {\bibfnamefont {R.}~\bibnamefont {Joyner}}, \
  and\ \bibinfo {author} {\bibfnamefont {W.~E.}\ \bibnamefont {Moerner}},\
  }\href {\doibase 10.1103/PhysRevE.91.062716} {\bibfield  {journal} {\bibinfo
  {journal} {Physical Review E}\ }\textbf {\bibinfo {volume} {91}},\ \bibinfo
  {pages} {062716} (\bibinfo {year} {2015})}\BibitemShut {NoStop}%
\bibitem [{\citenamefont {Calderon}(2016)}]{Calderon2016MotionTrajectory}%
  \BibitemOpen
  \bibfield  {author} {\bibinfo {author} {\bibfnamefont {C.~P.}\ \bibnamefont
  {Calderon}},\ }\href {\doibase 10.1103/PhysRevE.93.053303} {\bibfield
  {journal} {\bibinfo  {journal} {Physical Review E}\ }\textbf {\bibinfo
  {volume} {93}},\ \bibinfo {pages} {053303} (\bibinfo {year}
  {2016})}\BibitemShut {NoStop}%
\bibitem [{\citenamefont {Novotny}\ and\ \citenamefont
  {Hecht}(2012)}]{novotny2012principles}%
  \BibitemOpen
  \bibfield  {author} {\bibinfo {author} {\bibfnamefont {L.}~\bibnamefont
  {Novotny}}\ and\ \bibinfo {author} {\bibfnamefont {B.}~\bibnamefont
  {Hecht}},\ }\href@noop {} {\emph {\bibinfo {title} {Principles of
  Nano-Optics}}}\ (\bibinfo  {publisher} {Cambridge University Press,
  Cambridge, England},\ \bibinfo {year} {2012})\BibitemShut {NoStop}%
\bibitem [{\citenamefont {B{\"{o}}hmer}\ and\ \citenamefont
  {Enderlein}(2003)}]{Bohmer2003OrientationMicroscopy}%
  \BibitemOpen
  \bibfield  {author} {\bibinfo {author} {\bibfnamefont {M.}~\bibnamefont
  {B{\"{o}}hmer}}\ and\ \bibinfo {author} {\bibfnamefont {J.}~\bibnamefont
  {Enderlein}},\ }\href {\doibase 10.1364/JOSAB.20.000554} {\bibfield
  {journal} {\bibinfo  {journal} {Journal of the Optical Society of America B}\
  }\textbf {\bibinfo {volume} {20}},\ \bibinfo {pages} {554} (\bibinfo {year}
  {2003})}\BibitemShut {NoStop}%
\bibitem [{\citenamefont {Lieb}\ \emph {et~al.}(2004)\citenamefont {Lieb},
  \citenamefont {Zavislan},\ and\ \citenamefont
  {Novotny}}]{Lieb2004Single-moleculeImaging}%
  \BibitemOpen
  \bibfield  {author} {\bibinfo {author} {\bibfnamefont {M.~A.}\ \bibnamefont
  {Lieb}}, \bibinfo {author} {\bibfnamefont {J.~M.}\ \bibnamefont {Zavislan}},
  \ and\ \bibinfo {author} {\bibfnamefont {L.}~\bibnamefont {Novotny}},\ }\href
  {\doibase 10.1364/JOSAB.21.001210} {\bibfield  {journal} {\bibinfo  {journal}
  {Journal of the Optical Society of America B}\ }\textbf {\bibinfo {volume}
  {21}},\ \bibinfo {pages} {1210} (\bibinfo {year} {2004})}\BibitemShut
  {NoStop}%
\bibitem [{\citenamefont {Axelrod}(2012)}]{Axelrod2012FluorescenceStudy}%
  \BibitemOpen
  \bibfield  {author} {\bibinfo {author} {\bibfnamefont {D.}~\bibnamefont
  {Axelrod}},\ }\href {\doibase 10.1111/j.1365-2818.2012.03625.x} {\bibfield
  {journal} {\bibinfo  {journal} {Journal of Microscopy}\ }\textbf {\bibinfo
  {volume} {247}},\ \bibinfo {pages} {147} (\bibinfo {year}
  {2012})}\BibitemShut {NoStop}%
\bibitem [{\citenamefont {Backer}\ and\ \citenamefont
  {Moerner}(2014)}]{Backer2014ExtendingProcessing}%
  \BibitemOpen
  \bibfield  {author} {\bibinfo {author} {\bibfnamefont {A.~S.}\ \bibnamefont
  {Backer}}\ and\ \bibinfo {author} {\bibfnamefont {W.~E.}\ \bibnamefont
  {Moerner}},\ }\href {\doibase 10.1021/jp501778z} {\bibfield  {journal}
  {\bibinfo  {journal} {The Journal of Physical Chemistry B}\ }\textbf
  {\bibinfo {volume} {118}},\ \bibinfo {pages} {8313} (\bibinfo {year}
  {2014})}\BibitemShut {NoStop}%
\bibitem [{\citenamefont {Backer}\ and\ \citenamefont
  {Moerner}(2015)}]{Backer2015DeterminingStudy}%
  \BibitemOpen
  \bibfield  {author} {\bibinfo {author} {\bibfnamefont {A.~S.}\ \bibnamefont
  {Backer}}\ and\ \bibinfo {author} {\bibfnamefont {W.~E.}\ \bibnamefont
  {Moerner}},\ }\href {\doibase 10.1364/OE.23.004255} {\bibfield  {journal}
  {\bibinfo  {journal} {Optics Express}\ }\textbf {\bibinfo {volume} {23}},\
  \bibinfo {pages} {4255} (\bibinfo {year} {2015})}\BibitemShut {NoStop}%
\bibitem [{SI()}]{SI}%
  \BibitemOpen
  \href@noop {} {}\bibinfo {note} {See supplemental material for additional
  figures and derivation, which includes Refs.
  \cite{hell1994breaking,Rust2006Sub-diffraction-limitSTORM,Betzig2006ImagingResolution,Hess2006Ultra-HighMicroscopy,bouhelier2006field}}\BibitemShut
  {NoStop}%
\bibitem [{\citenamefont {Valeur}(2002)}]{valeur2012molecular}%
  \BibitemOpen
  \bibfield  {author} {\bibinfo {author} {\bibfnamefont {B.}~\bibnamefont
  {Valeur}},\ }\href@noop {} {\emph {\bibinfo {title} {Molecular Fluorescence:
  Principles and Applications}}}\ (\bibinfo  {publisher} {Wiley-VCH, Weinheim,
  Germany},\ \bibinfo {year} {2002})\BibitemShut {NoStop}%
\bibitem [{\citenamefont {Moon}\ and\ \citenamefont
  {Stirling}(2000)}]{moon2000mathematical}%
  \BibitemOpen
  \bibfield  {author} {\bibinfo {author} {\bibfnamefont {T.~K.}\ \bibnamefont
  {Moon}}\ and\ \bibinfo {author} {\bibfnamefont {W.~C.}\ \bibnamefont
  {Stirling}},\ }\href@noop {} {\emph {\bibinfo {title} {Mathematical Methods
  and Algorithms for Signal Processing}}}\ (\bibinfo  {publisher} {Prentice
  Hall, New Jersey},\ \bibinfo {year} {2000})\BibitemShut {NoStop}%
\bibitem [{\citenamefont {Nuttall}(1975)}]{Nuttall1975SomeCorresp.}%
  \BibitemOpen
  \bibfield  {author} {\bibinfo {author} {\bibfnamefont {A.}~\bibnamefont
  {Nuttall}},\ }\href {\doibase 10.1109/TIT.1975.1055327} {\bibfield  {journal}
  {\bibinfo  {journal} {IEEE Transactions on Information Theory}\ }\textbf
  {\bibinfo {volume} {21}},\ \bibinfo {pages} {95} (\bibinfo {year}
  {1975})}\BibitemShut {NoStop}%
\bibitem [{\citenamefont {Zhang}\ \emph {et~al.}(2018)\citenamefont {Zhang},
  \citenamefont {Lu}, \citenamefont {Ding},\ and\ \citenamefont
  {Lew}}]{Zhang2018ImagingFunction}%
  \BibitemOpen
  \bibfield  {author} {\bibinfo {author} {\bibfnamefont {O.}~\bibnamefont
  {Zhang}}, \bibinfo {author} {\bibfnamefont {J.}~\bibnamefont {Lu}}, \bibinfo
  {author} {\bibfnamefont {T.}~\bibnamefont {Ding}}, \ and\ \bibinfo {author}
  {\bibfnamefont {M.~D.}\ \bibnamefont {Lew}},\ }\href {\doibase
  10.1063/1.5031759} {\bibfield  {journal} {\bibinfo  {journal} {Applied
  Physics Letters}\ }\textbf {\bibinfo {volume} {113}},\ \bibinfo {pages}
  {031103} (\bibinfo {year} {2018})}\BibitemShut {NoStop}%
\bibitem [{\citenamefont {Mortensen}\ \emph {et~al.}(2010)\citenamefont
  {Mortensen}, \citenamefont {Churchman}, \citenamefont {Spudich},\ and\
  \citenamefont {Flyvbjerg}}]{Mortensen2010OptimizedMicroscopy}%
  \BibitemOpen
  \bibfield  {author} {\bibinfo {author} {\bibfnamefont {K.~I.}\ \bibnamefont
  {Mortensen}}, \bibinfo {author} {\bibfnamefont {L.~S.}\ \bibnamefont
  {Churchman}}, \bibinfo {author} {\bibfnamefont {J.~A.}\ \bibnamefont
  {Spudich}}, \ and\ \bibinfo {author} {\bibfnamefont {H.}~\bibnamefont
  {Flyvbjerg}},\ }\href {\doibase 10.1038/nmeth.1447} {\bibfield  {journal}
  {\bibinfo  {journal} {Nature Methods}\ }\textbf {\bibinfo {volume} {7}},\
  \bibinfo {pages} {377} (\bibinfo {year} {2010})}\BibitemShut {NoStop}%
\bibitem [{\citenamefont {Motegi}\ \emph {et~al.}(2013)\citenamefont {Motegi},
  \citenamefont {Nabika},\ and\ \citenamefont
  {Murakoshi}}]{Motegi2013Single-moleculeBilayers}%
  \BibitemOpen
  \bibfield  {author} {\bibinfo {author} {\bibfnamefont {T.}~\bibnamefont
  {Motegi}}, \bibinfo {author} {\bibfnamefont {H.}~\bibnamefont {Nabika}}, \
  and\ \bibinfo {author} {\bibfnamefont {K.}~\bibnamefont {Murakoshi}},\ }\href
  {\doibase 10.1039/c3cp51585k} {\bibfield  {journal} {\bibinfo  {journal}
  {Physical Chemistry Chemical Physics}\ }\textbf {\bibinfo {volume} {15}},\
  \bibinfo {pages} {12895} (\bibinfo {year} {2013})}\BibitemShut {NoStop}%
\bibitem [{\citenamefont {Lakowicz}(2006)}]{lakowicz2004principles}%
  \BibitemOpen
  \bibfield  {author} {\bibinfo {author} {\bibfnamefont {J.~R.}\ \bibnamefont
  {Lakowicz}},\ }\href@noop {} {\emph {\bibinfo {title} {Principles of
  Fluorescence Spectroscopy}}}\ (\bibinfo  {publisher} {Springer
  Science+Business Media, New York},\ \bibinfo {year} {2006})\BibitemShut
  {NoStop}%
\bibitem [{\citenamefont {Hell}\ and\ \citenamefont
  {Wichmann}(1994)}]{hell1994breaking}%
  \BibitemOpen
  \bibfield  {author} {\bibinfo {author} {\bibfnamefont {S.~W.}\ \bibnamefont
  {Hell}}\ and\ \bibinfo {author} {\bibfnamefont {J.}~\bibnamefont
  {Wichmann}},\ }\href@noop {} {\bibfield  {journal} {\bibinfo  {journal}
  {Optics letters}\ }\textbf {\bibinfo {volume} {19}},\ \bibinfo {pages} {780}
  (\bibinfo {year} {1994})}\BibitemShut {NoStop}%
\bibitem [{\citenamefont {Rust}\ \emph {et~al.}(2006)\citenamefont {Rust},
  \citenamefont {Bates},\ and\ \citenamefont
  {Zhuang}}]{Rust2006Sub-diffraction-limitSTORM}%
  \BibitemOpen
  \bibfield  {author} {\bibinfo {author} {\bibfnamefont {M.~J.}\ \bibnamefont
  {Rust}}, \bibinfo {author} {\bibfnamefont {M.}~\bibnamefont {Bates}}, \ and\
  \bibinfo {author} {\bibfnamefont {X.}~\bibnamefont {Zhuang}},\ }\href
  {\doibase 10.1038/nmeth929} {\bibfield  {journal} {\bibinfo  {journal}
  {Nature Methods}\ }\textbf {\bibinfo {volume} {3}},\ \bibinfo {pages} {793}
  (\bibinfo {year} {2006})}\BibitemShut {NoStop}%
\bibitem [{\citenamefont {Betzig}\ \emph {et~al.}(2006)\citenamefont {Betzig},
  \citenamefont {Patterson}, \citenamefont {Sougrat}, \citenamefont
  {Lindwasser}, \citenamefont {Olenych}, \citenamefont {Bonifacino},
  \citenamefont {Davidson}, \citenamefont {Lippincott-Schwartz},\ and\
  \citenamefont {Hess}}]{Betzig2006ImagingResolution}%
  \BibitemOpen
  \bibfield  {author} {\bibinfo {author} {\bibfnamefont {E.}~\bibnamefont
  {Betzig}}, \bibinfo {author} {\bibfnamefont {G.~H.}\ \bibnamefont
  {Patterson}}, \bibinfo {author} {\bibfnamefont {R.}~\bibnamefont {Sougrat}},
  \bibinfo {author} {\bibfnamefont {O.~W.}\ \bibnamefont {Lindwasser}},
  \bibinfo {author} {\bibfnamefont {S.}~\bibnamefont {Olenych}}, \bibinfo
  {author} {\bibfnamefont {J.~S.}\ \bibnamefont {Bonifacino}}, \bibinfo
  {author} {\bibfnamefont {M.~W.}\ \bibnamefont {Davidson}}, \bibinfo {author}
  {\bibfnamefont {J.}~\bibnamefont {Lippincott-Schwartz}}, \ and\ \bibinfo
  {author} {\bibfnamefont {H.~F.}\ \bibnamefont {Hess}},\ }\href {\doibase
  10.1126/science.1127344} {\bibfield  {journal} {\bibinfo  {journal}
  {Science}\ }\textbf {\bibinfo {volume} {313}},\ \bibinfo {pages} {1642}
  (\bibinfo {year} {2006})}\BibitemShut {NoStop}%
\bibitem [{\citenamefont {Hess}\ \emph {et~al.}(2006)\citenamefont {Hess},
  \citenamefont {Girirajan},\ and\ \citenamefont
  {Mason}}]{Hess2006Ultra-HighMicroscopy}%
  \BibitemOpen
  \bibfield  {author} {\bibinfo {author} {\bibfnamefont {S.~T.}\ \bibnamefont
  {Hess}}, \bibinfo {author} {\bibfnamefont {T.~P.}\ \bibnamefont {Girirajan}},
  \ and\ \bibinfo {author} {\bibfnamefont {M.~D.}\ \bibnamefont {Mason}},\
  }\href {\doibase 10.1529/biophysj.106.091116} {\bibfield  {journal} {\bibinfo
   {journal} {Biophysical Journal}\ }\textbf {\bibinfo {volume} {91}},\
  \bibinfo {pages} {4258} (\bibinfo {year} {2006})}\BibitemShut {NoStop}%
\bibitem [{\citenamefont {Bouhelier}(2006)}]{bouhelier2006field}%
  \BibitemOpen
  \bibfield  {author} {\bibinfo {author} {\bibfnamefont {A.}~\bibnamefont
  {Bouhelier}},\ }\href@noop {} {\bibfield  {journal} {\bibinfo  {journal}
  {Microscopy research and technique}\ }\textbf {\bibinfo {volume} {69}},\
  \bibinfo {pages} {563} (\bibinfo {year} {2006})}\BibitemShut {NoStop}%
\end{thebibliography}%

\end{document}